\documentclass[
 twocolumn,superscriptaddress,aps,amsmath,amssymb,floatfix]{revtex4-2}

\usepackage{graphicx}
\usepackage{dcolumn}
\usepackage{bm}
\usepackage[breaklinks=true,colorlinks,citecolor=blue,linkcolor=blue,urlcolor=blue]{hyperref}

\usepackage{physics}
\usepackage[normalem]{ulem}

\newcommand{\sz}{\hat \sigma_z}
\newcommand{\sy}{\hat \sigma_y}
\newcommand{\sx}{\hat \sigma_x}

\newcommand{\aop}{\hat a}
\newcommand{\adop}{\hat a ^\dagger}

\newcommand{\figpanel}[2]{Fig.~\hyperref[#1]{\ref*{#1}(#2)}}

\begin{document}

\title{Flying atom back-reaction and mechanically generated photons from vacuum}

\author{Alberto Mercurio}
\email{alberto.mercurio@unime.it}
\affiliation{%
 Dipartimento di Scienze Matematiche e Informatiche, Scienze Fisiche e  Scienze della Terra,
	Universit\`{a} di Messina, I-98166 Messina, Italy
}%

\author{Simone De Liberato}
\affiliation{%
 School of Physics and Astronomy, University of Southampton, Southampton, SO17 1BJ, United Kingdom
}%

\author{Franco Nori}
\affiliation{Theoretical Quantum Physics Laboratory, Cluster for Pioneering Research, RIKEN, Saitama 351-0198, Japan}
\affiliation{Center for Quantum Computing, RIKEN, Wako-shi, Saitama 351-0198, Japan}
\affiliation{Physics Department, The University of Michigan, Ann Arbor, Michigan 48109-1040, USA.}

\author{Salvatore Savasta}
\email{salvatore.savasta@unime.it}
\affiliation{%
 Dipartimento di Scienze Matematiche e Informatiche, Scienze Fisiche e  Scienze della Terra,
	Universit\`{a} di Messina, I-98166 Messina, Italy
}%

\author{Roberto Stassi}
\affiliation{%
 Dipartimento di Scienze Matematiche e Informatiche, Scienze Fisiche e  Scienze della Terra,
	Universit\`{a} di Messina, I-98166 Messina, Italy
}%
\affiliation{Center for Quantum Computing, RIKEN, Wako-shi, Saitama 351-0198, Japan}

\date{\today}

\begin{abstract}
We investigate the dynamics of a two-level atom flying through a photonic cavity when the light-matter interaction is in the ultrastrong coupling regime. We adopt a closed full quantum description that takes into account the quantization of the atom center-of-mass motion in addition to its internal degree of freedom and to the quantized photonic cavity field. We find that multiple qualitatively different dynamical regimes are achievable according to two key figures of merit: the ratio between the kinetic energy and the bare excitation energies, and the product of these bare energies with the time the atom takes to fly through the cavity.
According to the values of those figures of merit, the atom can be reflected by the \emph{dressed} vacuum, or can convert part of its kinetic energy into real excitations which might be emitted out of the cavity. In the first case, the atom experiences a \emph{quantum regenerative braking} mechanism, based on temporary storage of energy into virtual excitations.
\end{abstract}

\maketitle

In pioneering experiments, flying atoms passing through a cavity were used to investigate the light-matter strong coupling regime \cite{kaluzny1983,  meschede1985one, rempe1987observation, thompson1992}. These experiments allowed to test fundamental aspects of measurement theory \cite{braginskii1981non, brune2008process}, and led to proof-of-concept demonstrations of basic steps in quantum information processing \cite{raimond2001, haroche2013}.

Flying atoms in a cavity have also been proposed to explore vacuum emission phenomena.   
For example, a theoretical study \cite{scully2003enhancing} has shown that ground-state atoms, if accelerated through a high-Q microwave cavity, can generate radiation with an intensity which can exceed that of the Unruh acceleration radiation in free space by many orders of magnitude.

Several other theoretical works predicted the generation of particles from the vacuum upon dynamical modulation of the cavity system: varying in time the boundary condition of an electromagnetic field (dynamical Casimir effect) \cite{fulling1976radiation, johansson2009dynamical, wilson2011observation},  moving a qubit inside a single-mode cavity \cite{agusti2021qubit}, having electrical currents transit through a photonic resonator \cite{cirio2016ground,cirio2019multielectron}, or an atom in systems with broken Lorentz invariance \cite{svidzinsky2019excitation}.

In the ultrastrong coupling (USC) regime, in which the light-matter interaction strength is comparable or larger than the bare  frequencies of the matter and/or light resonances, vacuum radiation has also been predicted to be emitted when the atom-light interaction is non-adiabatically modulated \cite{de2007quantum,de2009extracavity} or switched on or off \cite{garziano2013switching,stassi2013,di2017feynman}. 

Since its first experimental observation in doped quantum wells \cite{anappara2009signatures}, experimental progress in cavity quantum electrodynamics led to the achievement of USC in many different systems and with ever-increasing coupling strengths
\cite{gunter2009sub,niemczyk2010circuit,scalari2012ultrastrong, gambino2014exploring,goryachev2014high, pirkkalainen2015cavity, baust2016ultrastrong, benz2016single, george2016multiple, forn2017ultrastrong, yoshihara2017superconducting, yoshihara2017characteristic, bayer2017terahertz, barachati2018tunable, yoshihara2018inversion, shi2018ultrastrong, flower2019experimental,rajabali2021polaritonic,kockum2019ultrastrong, forn2019ultrastrong}.
This in turn stimulated an intense theoretical effort  \cite{kockum2019ultrastrong, forn2019ultrastrong,ma2015three, kockum2017deterministic, kockum2017frequency, stassi2017quantum, macri2018nonperturbative} and
a rich phenomenology has been predicted to occur in such a non-perturbative coupling regime. For example, a single photon can excite multiple atoms simultaneously \cite{garziano2016one, macri2020spin}, lasing emission is modified \cite{bamba2016laser},
mechanical resonators can interact through virtual photons \cite{distefano2019interaction},
electronic ground-state configurations can be modified \cite{wang2021theoretical},
and Higgs-like phenomena can be observed breaking the parity symmetry of the system \cite{garziano2014vacuum, wang2022detecting}. The USC regime could be also applied in quantum technologies \cite{stassi2020scalable}.

\begin{figure}[b]
    \centering
    \includegraphics[width = \linewidth]{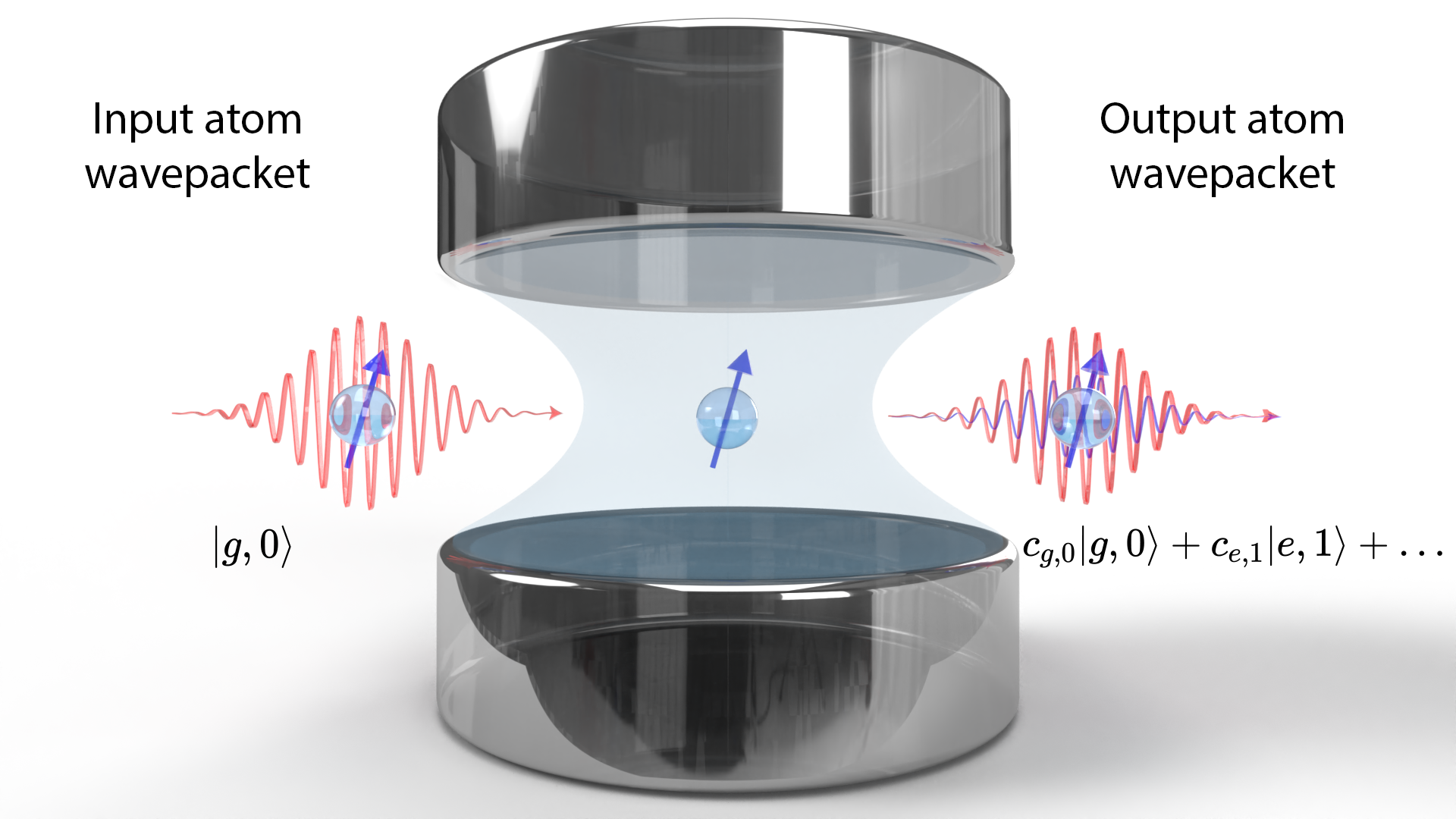}
    \caption{Representation of an atom wavepacket passing through a cavity resonator. Depending on its kinetic energy, the atom can leave the cavity entangled with a photon mechanically generated from the vacuum.}
    \label{fig:pictorial representation}
\end{figure}

In this article, we investigate the effects caused by an atomic two-level system flying through a cavity, where the light-matter interaction is in the USC regime. We consider the system initially prepared in its lowest internal energy state (zero atomic and light excitations) with the atom flying toward the cavity, and we calculate the quantum dynamics of the coupled system taking into account the interaction between center-of-mass and internal degrees of freedom of the atom.
Such an interaction is due to the light-matter interaction changing the ground state energy inside the cavity, which also determines an effective potential barrier which can affect the center-of-mass motion.

Simulating the time evolution of the coupled system as the atom passes through the cavity, we find qualitatively different behaviours depending on two figures of merit. The first is the initial kinetic energy of the atom when compared to the height  of the barrier.
If the initial kinetic energy is much lower than the barrier, the atom slows down and gets reflected, with at least part of the kinetic energy temporarily stored in virtual excitations of the atom-photon system. If the kinetic energy is higher than the barrier, the atom passes instead through the cavity. 
The second figure of merit is instead the product between the bare excitation frequencies of the system and the time the atom takes to traverse the cavity. If this quantity is small, the atom experiences a non-adiabatic modification in its interaction with the cavity field.
Then the dynamics becomes inelastic, and part of the kinetic energy is converted into
atomic and photonic real excitations entangled with the atomic center-of-mass.

The experimental observation of the physics described in our theoretical proposal requires
to achieve the USC regime with single dipoles. While most systems described in the introduction rely on collective coupling to reach such a regime, there has been a sustained effort to reduce the number of dipoles   \cite{ballarini2019polaritonics}.
Moreover, single molecule USC has been  achieved using vibrational degrees of freedom in plasmonic picocavities \cite{benz2016single}. Standard electronic strong coupling has also been achieved in single molecules trapped in plasmonic nanogaps \cite{chikkaraddy2016single-molecule_2016}, with the possibility to reach USC theoretically predicted \cite{kuisma2022ultrastrong}.



\vspace{4mm}
\noindent \textbf{Results}

\noindent The atom-cavity system is governed by a time-independent Hamiltonian, that reads ($\hbar = 1$) 
\begin{equation}
    \label{eq: total Hamiltonian (coulomb)}
    \hat{\mathcal{H}}^{\rm (\mathfrak{g})} = \frac{\hat{p}^2}{2 m} + \mathcal{\hat{H}}_{\rm R}^{(\rm \mathfrak{g} )}~,
\end{equation}
where $\mathfrak{g}=\{\rm{c,d}\}$ refers to the chosen gauge. We explicitly indicate the gauge because, as better explained in the Methods section, gauge invariance can cause non-trivial problems in the USC regime, and different aspects of the physics are better understood in one gauge or the other \cite{debernardis2018breakdown,distefano2019resolution,stokes2021ultrastrong,settineri2021gauge}. In the main body of the article we choose to use the Coulomb gauge (c), while the comparisons with the dipole gauge (d) can be found in the Methods section.
The first term of the r.h.s. of Eq.\,\eqref{eq: total Hamiltonian (coulomb)} is the 1D kinetic energy of the atom center-of-mass, where $\hat p = - i d/dx$ is the linear momentum and $m$ is the mass. The second term, $\mathcal{\hat{H}}_{\rm R}^{(\rm c )}$, is instead a quantum Rabi Hamiltonian describing the interaction between the atom and a single-mode cavity in the Coulomb gauge, which is derived applying the generalized minimal coupling replacement \cite{distefano2019resolution},
\begin{eqnarray}
    \label{eq: rabi hamiltonian (coulomb)}
    \mathcal{\hat{H}}_{\rm R}^{(\rm c )} &=& \omega_c \adop \aop + \frac{1}{2} \omega_a \left\{ \sz \cos \left[ 2 \eta (\hat{x} ) ( \aop + \adop ) \right] \right. \nonumber \\
    &+& \left. \sy \sin \left[2 \eta (\hat{x} ) (\aop + \adop ) \right] \right\}~.
\end{eqnarray}
Here $\omega_c$ and $\omega_a$ are, respectively, the cavity and atom frequencies, $\aop$ ($\adop$) is the annihilation (creation) operator for the single-mode cavity field, and $\hat{\sigma}_i$ ($i = x, y, z$) are Pauli matrices operating on the internal states of the two-level atom  $\{\ket{g},\ket{e}\}$.  

For the sake of definiteness, we assume that the normalized space-dependent light-matter coupling strength has a Gaussian shape centered in the cavity center with width $\mu_c$ and maximal intensity $\eta_0$ \cite{brune1994lamb, brune1996quantum}
\begin{align}
\eta(x) = \eta_0 \exp \left[- \frac{x^2}{ 2 \mu_c^2} \right].
\end{align}
This assumption is not essential and different field profiles could be considered.


In the USC regime, the eigenstates of the Hamiltonian in Eq.\,\eqref{eq: rabi hamiltonian (coulomb)} contain virtual excitations if represented in the bare atom and photon basis, with the $l$th state for the atom localised at position $x$ having the form 
\begin{eqnarray}
\ket{ l}_x=\sum_{n=0}^{\infty} {c_{ l}^{(2n)}(x)\ket{g, 2n}+c_{l}^{(2n+1)}(x)\ket{e, 2n+1}}, 
\end{eqnarray}
where the states $\ket{g,n}$ and $\ket{e,n}$ form a basis of the quantum Rabi Hamiltonian, describing the atom in the ground or excited state with $n$ photons in the cavity.
The exact form of the coefficients $c_{l}^{(i)}(x)$ depend on the coupling strength at position $x$, $\eta(x)$, and on the chosen gauge. 

\noindent \textbf{Dynamics.}~ We numerically study the full quantum dynamics of the system applying the time evolution operator, $\hat T^{(c)}={\rm exp}[-i \hat H^{(c)} t]$, to the initial state. In our simulations we do not consider dissipation, which has been shown to have only a limited impact on the population of virtual excitations \cite{de2017virtual}.
At the initial time $t_0=0$, the system's state is assumed to be in the factorised form
\begin{eqnarray}
\label{eq:initial} 
\ket{\Psi (t_0)}= \ket{\varphi(t_0)} \ket{g, 0}.
\end{eqnarray} 
The center-of-mass initial wave-function, centered at $x_0$ and of width $\mu_s$  is chosen as
\begin{align}
\bra{x} \ket{\varphi(t_0)} = G \left( \frac{x-x_0}{\mu_s} \right) \exp(i k_0 x), 
\end{align}
where $G(x)$ is a normalised symmetric Gaussian function of unit variance.
Initially, the atom is placed outside the cavity, $\bra{\varphi(t_0)} \hat{x} \ket{\varphi(t_0)} = x_0$, with $|x_0|\gg \mu_c$, implying $\eta(x_0)\approx 0$, and
the initial momentum is directed towards the cavity, $\bra{\varphi(t_0)} \hat{p} \ket{\varphi(t_0)} = k_0$, corresponding to an initial velocity $v_0=k_0/m$.

\begin{figure}[t]
    \centering
    \includegraphics[width = \linewidth]{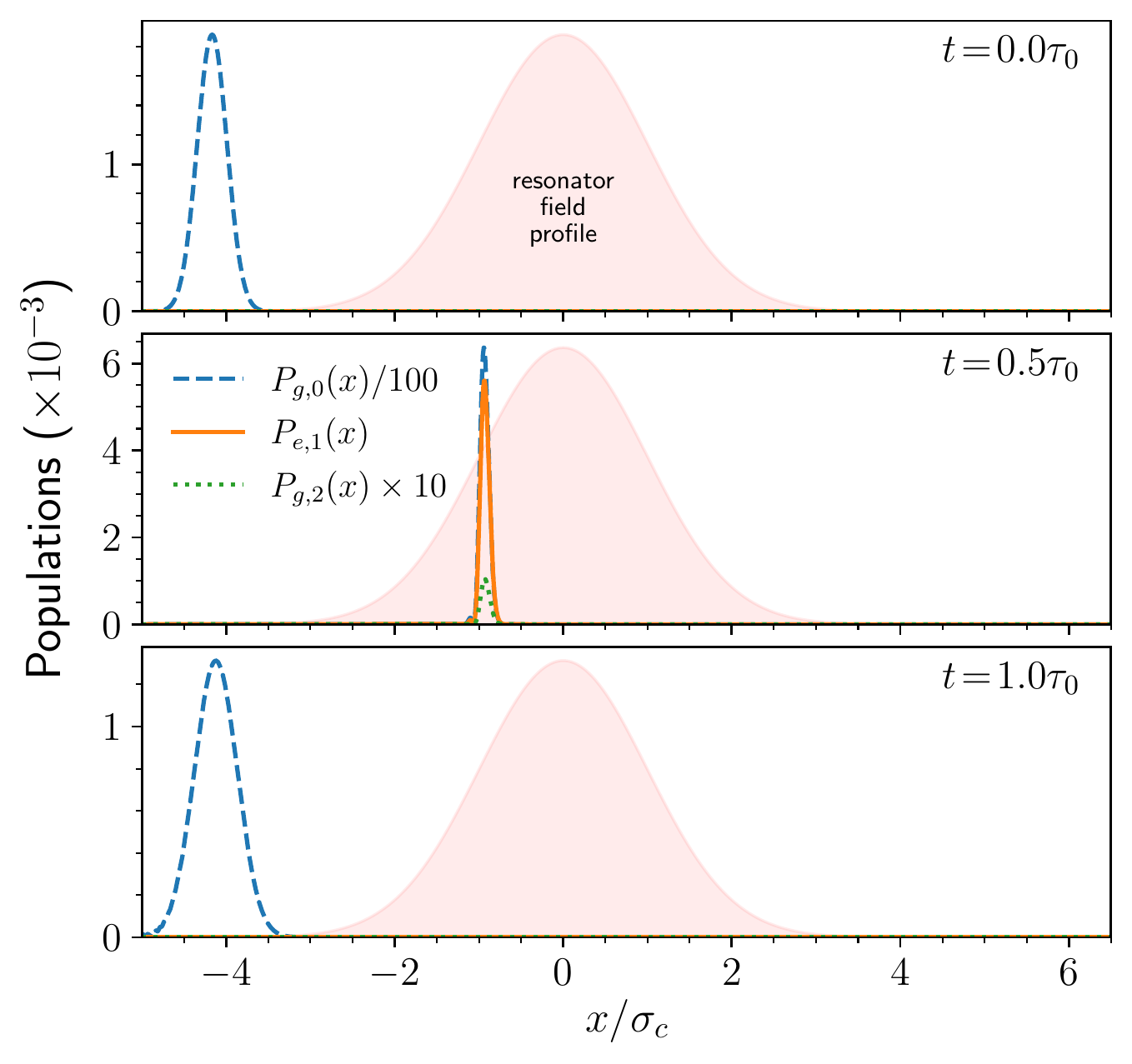}
    \caption{Snapshots of the Gaussian wave-packets associated to the states $\ket{g,0}, \ket{e,1}$ and $\ket{g,2}$,  when the initial normalized kinetic energy is ${E}_K = 0.02$. At $t=0\,\tau_0$, the atom and the cavity are both in their ground state. At $t=0.5\,\tau_0$, when the atom is in the cavity, the atom-light excitations increase, while at the same time the atom's kinetic energy decreases. At $t=\,\tau_0$, the atom leaves the cavity with conserved initial kinetic energy. Here, $\omega_a=\omega_c$, $\eta_0 = 0.3$, $\lambda_0 \equiv 2 \pi / k_0 = \mu_c / 10$.
    }
    \label{fig: snapshots reflection}
\end{figure}

In the USC coupling, the local vacuum energy depends on the strength of the light-matter coupling. This can be better understood in the dipole gauge (see Methods), where the quantum Rabi Hamiltonian has a term equal to $\omega_c \eta^2(\hat{x})$ \cite{distefano2019resolution, mercurio2022regimes}. In the case where the atom is considered to be fixed inside the cavity, this contribution introduces only an energy shift, and can often be neglected. However, when considering a space-dependent coupling strength $\eta(x)$ \cite{Ian2012Excitation}, this term corresponds to a space-dependent effective potential which generates a force affecting the atom dynamics. Working in the USC regime (in which we expect the maximum value of the normalized coupling $\eta_0$ to be of the order of $0.1$ or more), a meaningful figure of merit to study the mechanical motion of the atom is therefore the initial kinetic energy normalised over the photon frequency
\begin{align}
    {E}_K= \frac{mv_0^2}{2\omega_c}.
\end{align}
Moreover, we expect that the atom-cavity system remains in its ground state only when the evolution is adiabatic, which in the present system should mean that the time the atom takes to see an increase of $\eta(x)$ from $0$ to $\eta_0$, is substantially longer than the optical period of the system. In the quasi-resonant case $\omega_a\approx\omega_c$, which we will consider in the rest of this paper, we can then introduce the adiabatic figure of merit 
\begin{align}
\label{Xi}
\Xi=\frac{v_0}{\omega_c\mu_c},    
\end{align}
and expect the evolution to be adiabatic when $\Xi\ll 1$.

\begin{figure}[b]
    \centering
    \includegraphics[width = \linewidth]{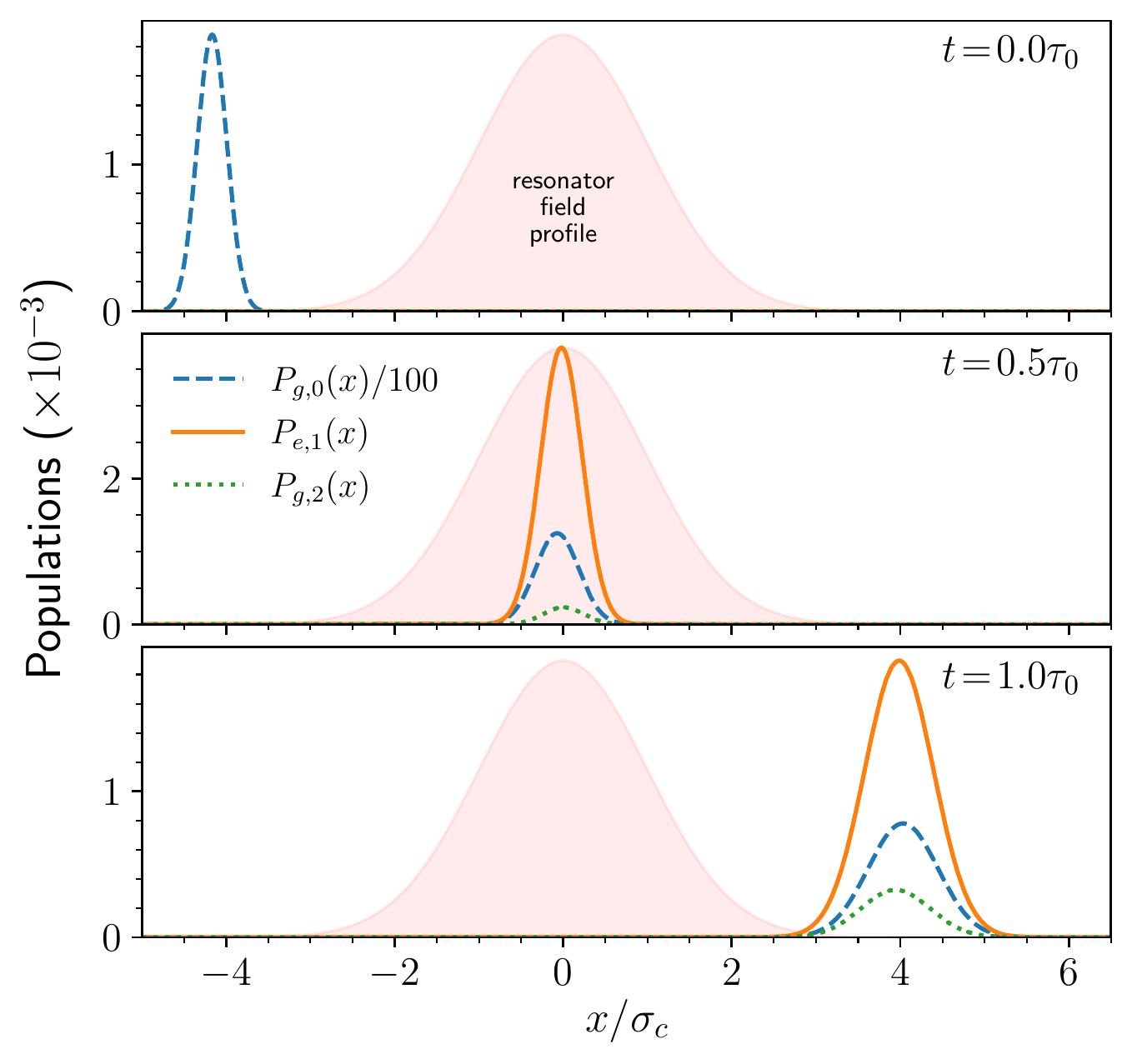}
    \caption{Snapshots of the Gaussian wave packets associated to the states $\ket{g,0}, \ket{e,1}$ and $\ket{g,2}$,  when the initial normalized kinetic energy is $E_k = 40$. At $t=0\,\tau_0$, the atom and the cavity are both in their ground state. At $t=0.5\,\tau_0$, when the atom enters the cavity, part of the kinetic energy is converted into atom-light excitations. At $t=\,\tau_0$, the atom leaves the cavity in a superposition of light-matter excited states. Here, $\eta_0 = 0.3$.}
    \label{fig: snapshots transmission}
\end{figure}

We start our simulation choosing two different values of the normalised kinetic energies $E_K=0.02$ and $E_K=40$, which will allow us to explore two completely different dynamical regimes. We will measure the time in units of $\tau_0=\abs{2x_0/v_0}$, which is the time the atom would take to fly unimpeded through the cavity to a position symmetric to the initial one with respect to the cavity center. In all the simulations we will fix the following parameters: $\omega_a=\omega_c$, $\eta_0 = 0.3$, $x_0=-(25 / 6)\mu_c$,
and $k_0 = 10\times 2\pi/\mu_c$. The mass $m$ is varied to change the initial kinetic energy. 

\begin{figure}[b]
    \centering
    \includegraphics[width = \linewidth]{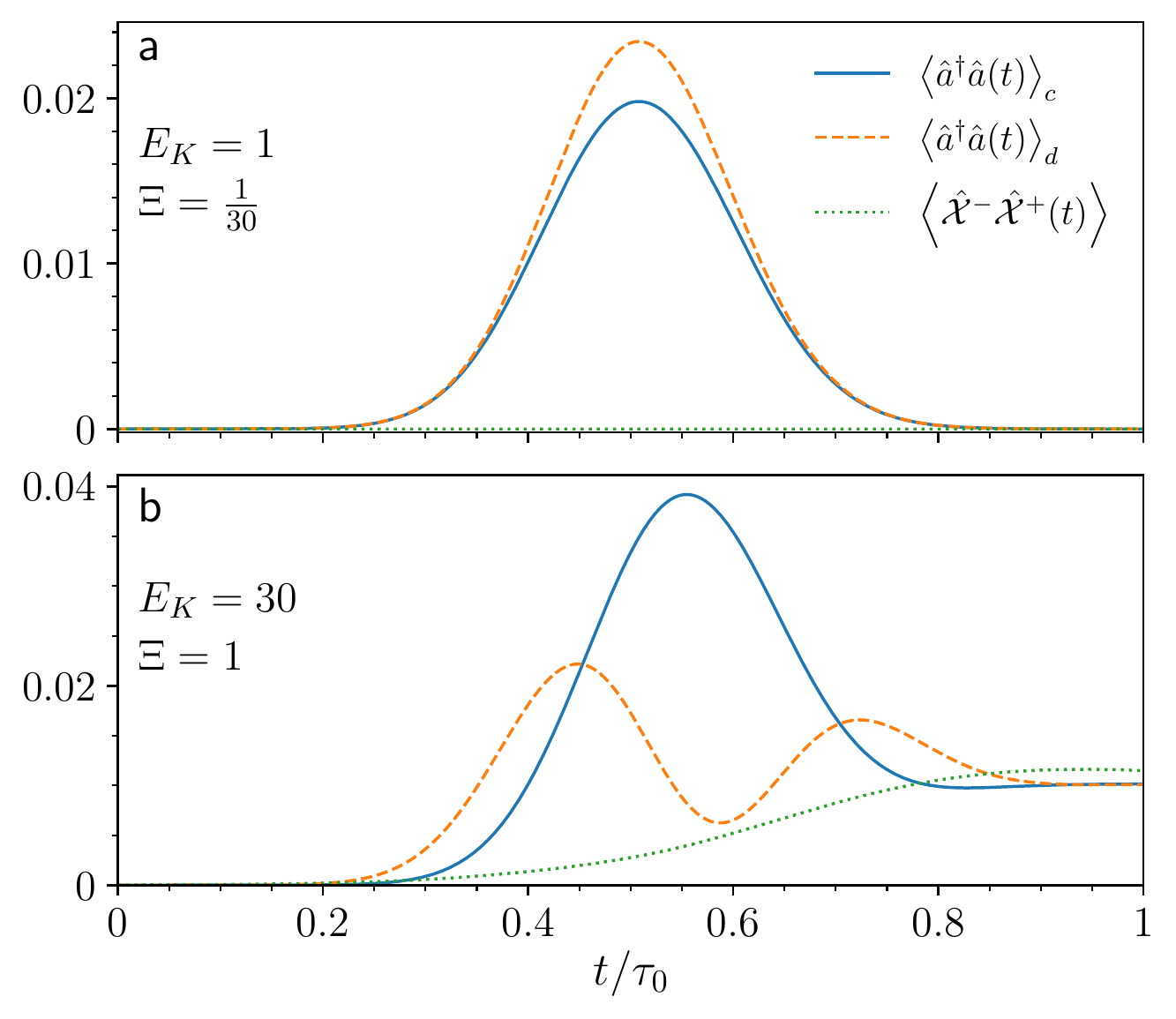}
    \caption{Comparison of the time evolution of the photon rates predicted by the standard expectation values $\langle \hat a^\dagger\hat a\rangle$ in the Coulomb (blue solid curve) and dipole gauge (orange dashed curve), and the one calculated making use of the expectation value $\langle \hat{\mathcal{X}}^{-} \hat{\mathcal{X}}^{+}\rangle$ (dotted green curve) during the atom flight. \textbf{a} The initial normalised kinetic energy of the atom is  ${E}_K  = 1$. The dynamics is adiabatic and no photons are generated; in fact $\langle \hat{\mathcal{X}}^{-} \hat{\mathcal{X}}^{+}\rangle$ is zero along the atom trajectory. \textbf{b} The initial kinetic energy is ${E}_K = 30$. In this case the process is non-adiabatic and real photons are generated.}
    \label{fig: gauge comparison}
\end{figure}

In Figure \ref{fig: snapshots reflection}, we show three snapshots of the wave-packet evolution for the initial normalised kinetic energy ${E}_K = 0.02$, corresponding to $\Xi\approx 6.6\times 10^{-4}$. At $t = 0$ the wave packet is outside the cavity, and it moves from the left to the right with initial velocity $v_0$. Before the atom enters the cavity, it remains in the initial state $\ket{g, 0}$, since $\bra{\varphi(t)} \eta(\hat{x})\ket{\varphi(t)} \approx 0$.
When the atom enters the cavity, as $\Xi\ll 1$, it sees the ground-state energy change adiabatically, and it thus remains approximately in its local ground state $\ket{0}_x$. We observe that, approaching the cavity center, the populations of virtual excitations forming the ground state (i.e., $\ket{e, 1}$ and $\ket{g, 2}$) increase, meanwhile the group velocity of the wave packet decreases (not shown in Fig.~\ref{fig: snapshots reflection}). At $t \simeq \tau_0 / 2$ the atom stops and reverses its motion. The atom then moves backwards from right to left leaving the cavity with momentum opposite of the initial one. At $t \simeq \tau_0$ the atom has left the cavity with the  kinetic energy and absolute value of the linear momentum essentially coincident with the initial ones.
The reflection happens because the gradient of the vacuum energy as a function of the atom position is positive when the atom enters the cavity, resulting in a force pushing back the atom when the kinetic energy is lower than the effective potential barrier.
It can be shown that excitations in Fig.~\ref{fig: snapshots reflection} for $t = 0.5 \tau_0$ 
are virtual. A hint comes from the fact that essentially no excitation is present when the atom leaves the cavity ($t = \tau_0$). A more direct evidence will be provided later (see Fig.~\ref{fig: gauge comparison}a).

In Figure \ref{fig: snapshots transmission} we show the numerical results obtained for an initial kinetic energy ${E}_K = 40$, corresponding to $\Xi \approx 1.3$. Not only in this case the kinetic energy is much larger than the barrier height, allowing the atom to pass through the cavity, but now the atom feels a non-adiabatic change of the local coupling strength $\eta(x)$ entering the cavity. Consequently, the system's state does not adiabatically follow the local ground state and the atomic and photonic excited states become populated. 
In this case the interaction of the atom with the cavity converts a fraction of the atomic kinetic energy into real excitations: both atomic excitations and real photons can persist after the atom leaves ($\tau = \tau_0$) and eventually leaks out of the system.
\begin{figure}[b]
    \centering
    \includegraphics[width = \linewidth]{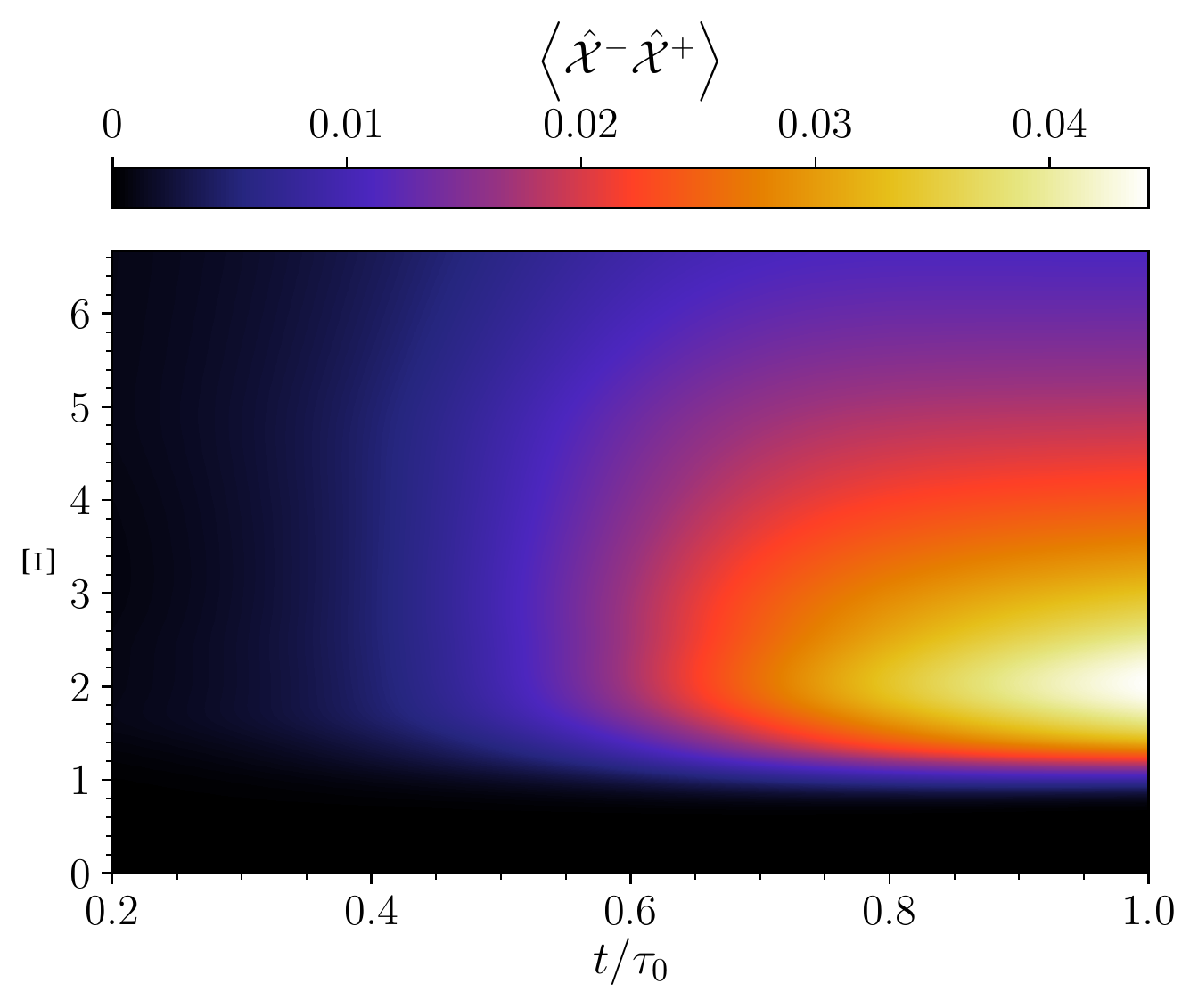}
    \caption{Rate of photons emitted as a function of time and the adiabatic parameter $\Xi$. At both small and large kinetic energies there is no creation of photons on the output, as explained in the text.}
    \label{fig: a_d a - E_k VS t}
\end{figure}

\noindent  \textbf{Emitted photon rates.}~ The rate of photon emission out of a cavity is usually assumed proportional to the mean cavity-photon number $\expval{\adop \aop}$. As better detailed in the Methods section, in the USC regime this ceases to be true
and the output has to be calculated using instead the quantity $\langle \hat{\mathcal{X}}^{-} \hat{\mathcal{X}}^{+}\rangle$  \cite{ridolfo2012photon, settineri2021gauge}, where the operator $\hat{\mathcal{X}}^{+} (\hat{\mathcal{X}}^{-})$ is related to the positive (negative) frequencies of the electric field operator, where 
\begin{align}
\hat{\mathcal{X}}^{+} = \sum_{l, k > l} i\mel{l}{ (\aop - \adop)}{k} \dyad{l}{k},    
\end{align}
and $\hat{\mathcal{X}}^{-} = ( \hat{\mathcal{X}}^{+} )^\dag$, with $\ket{l}$ the system eigenstates ordered for increasing values of the energy. However, when the atom is outside the cavity, the cavity-matter coupling is zero, and we can retain $\expval{\adop \aop}$. 
In Figure \ref{fig: a_d a - E_k VS t}, the mean  number of real photons $\langle \hat{\mathcal{X}}^{-} \hat{\mathcal{X}}^{+}\rangle$ is shown as a function of time for different values of the adiabatic parameter $\Xi$. 
We can clearly see that photons are emitted only when $\Xi>1$. As $\Xi$ increases, the emission slowly tapers off because the time the atom spends into the cavity grows shorter. Eventually, it becomes shorter than the Rabi oscillation period and the atom leaves the cavity before any meaningful interaction with the photonic field has the time to happen.
\begin{figure}[t]
    \centering
    \includegraphics[width = \linewidth]{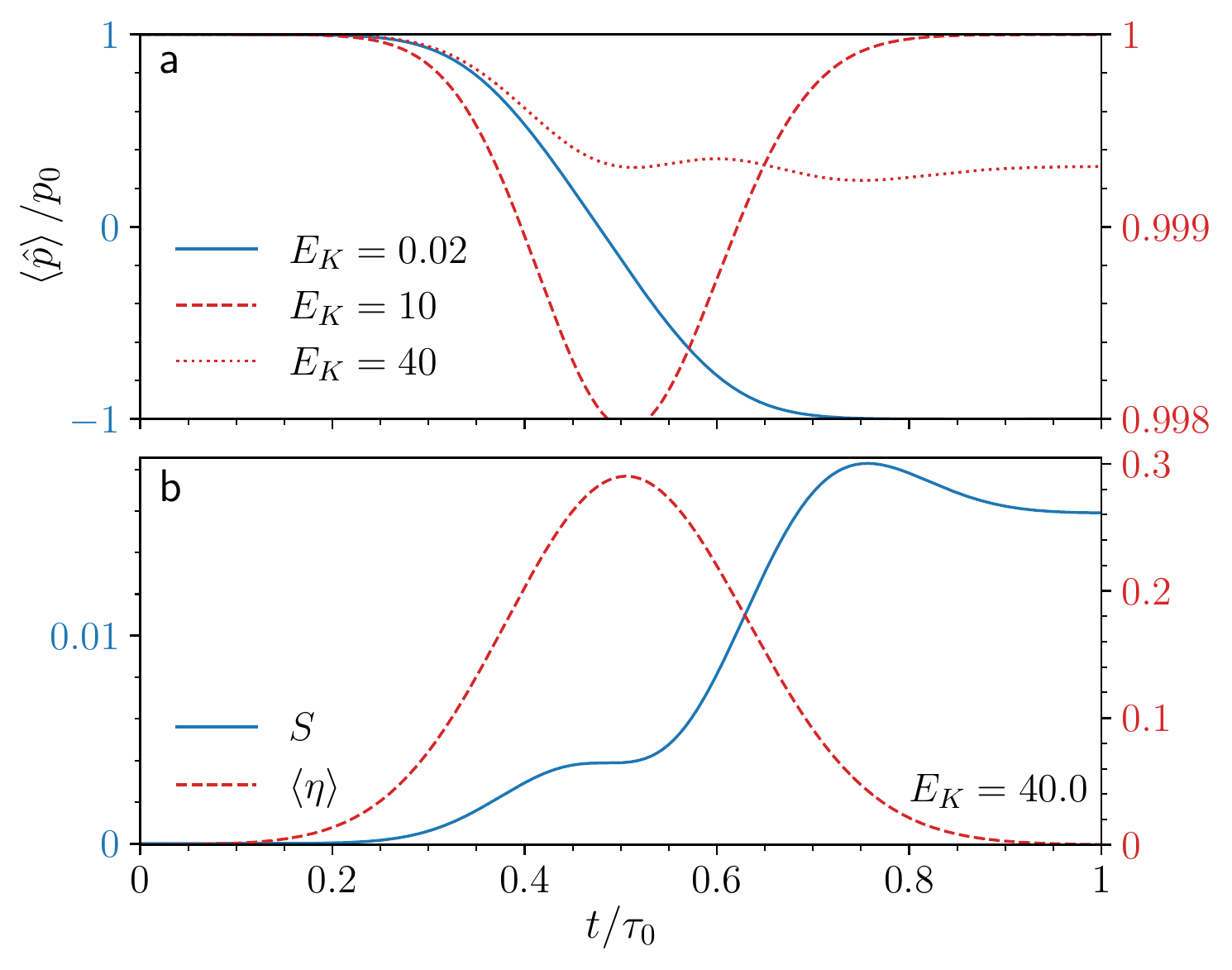}
    \caption{\textbf{a} Expectation value of the normalized atom momentum as a function of time for different initial kinetic energy values, ${E}_K = 0.02, 10, 40$. For ${E}_K = 0.02$, the atom is reflected and the momentum changes sign. For ${E}_K = 10$, the atom passes through the cavity, there is no creation of excitations, and the momentum comes back to its initial value. For ${E}_K = 0.02$, we can observe creation of photons and atom excitations, and the momentum accordingly decreases. \textbf{b} von-Neumann entropy as a function of time for an atom entering the cavity with initial kinetic energy ${E}_K = 40$. Other parameters are the same adopted in the other figures.}
    \label{fig: momentum and entanglement}
\end{figure}

\noindent  \textbf{Entanglement.}~ At every instant, the state of the system can be expressed as
\begin{equation}
    \label{eq: state bipartited decomposition}
    \ket{\Psi (t)} = \sum_{i} c_{i} (x,t) \ket{\varphi(t)} \ket{R_i}~,
\end{equation}
where $\ket{R_i}$ are states which span a basis of the quantum Rabi Hamiltonian (e.g., $\ket{g, 0}$, $\ket{g, 1}$, $\ket{e, 0},\ldots$). Although we chose an initially factorizable state in Eq.~\ref{eq:initial}, when the dynamics is non-adiabatic the atomic and photonic excitations generated in the process are entangled with the atom position.
In Figure \ref{fig: momentum and entanglement}b we plot the entanglement calculated using the von-Neumann entropy $S = - \Tr[\rho_{\rm R} \log_m (\rho_{\rm R})]$, where $\rho_{\rm R} = \Tr_\varphi [\rho]$ is the reduced density matrix obtained tracing out the atom center-of-mass motion. The coefficient $m$ is a cut-off to the dimension of the Hilbert space describing the atom and the cavity mode. In our simulation, we choose $m = 4$, since higher lying states have vanishing populations for the parameter range considered. 
 In Fig.\,\ref{fig: momentum and entanglement}b we see that, after the atom leaves the cavity, the entanglement remains constant.


\vspace{4mm}
\noindent \textbf{Discussion}

\noindent We investigated the dynamics of a flying atom prepared initially in its lowest internal energy state (e.g., the electronic ground state)   that passes through a cavity with zero photons and interacts with it in the USC regime. 
We were thus able to show how the generation of excitations out of the vacuum affects the center of mass dynamics. We also found that any temporary reduction of the atomic velocity due to a gradient of light-matter interaction strength results into a temporary energy storing in virtual excitations.

When the atom enters the cavity, it  experiences a force due to the gradient in the vacuum energy felt by the atom. This leads to a {\em quantum regenerative braking effect} in which the kinetic energy is temporarily stored in virtual atom-light excitations inside the cavity. Different cases can take place, according to whether the final momentum and energy of the atom are changed by the interaction.
If the kinetic energy is much lower than the difference between the ground state energies inside and outside the cavity, the atom is reflected. If instead its kinetic energy is large enough, the atom is transmitted.
The condition to create real excitations is connected to the adiabatic parameter $\Xi$. If it is small enough, the atom sees and adiabatic change in the light-matter interaction and in its ground state. In this case,  the interaction is elastic, and the atom's final kinetic energy is the same as the initial one. In the opposite case, the interaction is non-adiabatic and inelastic, as part of the initial kinetic energy is converted into real excitations.

These predictions might motivate new experimental ways to convert virtual photons into real ones. Our theoretical proposal could be experimentally observed in plasmonic picocavities or using single molecules trapped in plasmonic nanogaps. Nevertheless, the achievement of the USC with single molecule is still a challenging topic. Moreover, the experimental verification of this effect could verify also the signatures of the counter rotating terms, which are only present in the USC regime.

\vspace{4mm}
\noindent \textbf{Methods}

\noindent \textbf{Derivation of the quantum Rabi Hamiltonian.} The quantum Rabi Hamiltonian in Eq.~\ref{eq: rabi hamiltonian (coulomb)} approximates the physics of an atom coupled to a photonic resonator as a two-level system coupled to a single photonic mode. We are thus performing two key approximations: neglecting higher-lying matter states and photonic modes.

The first approximation leads to major problems when the coupling between light and matter enters in the USC regime, as gauge invariance breaks in the reduced Hilbert space and the two usually employed gauges, the Coulomb one and the dipolar one, provide completely different results, with the latter being the most correct one \cite{debernardis2018breakdown}.
Recently, a consistent description of the quantum Rabi model was introduced, which is able to provide gauge-invariant physical results in any interaction regime \cite{distefano2019resolution}. It has been shown that this description is closely connected to lattice gauge theories \cite{savasta2021gauge}. Moreover, it was shown in Ref. \cite{stokes2021ultrastrong} that adding a time dependent coupling may also lead to gauge-dependent predictions. These ambiguities have been solved in Ref. \cite{settineri2021gauge}. To derive the dynamics of our system, we take into account all these recent results regarding gauge invariance.

The neglect of higher photonic modes had also been recognised as potentially problematic for large values of the coupling strength \cite{sanchez2018resolution}, as it can lead to unphysical superluminal signalling. Still, the main issue for the present work regards the calculation of the energy barrier between the vacua outside and inside the cavity. A multimode exact calculation would have to consider that the cavity is effective at confining the electromagntic radiation up to a cutoff frequency $\omega_M$, and the energy difference felt by the atom inside and outside the cavity should then be calculated as a difference between a sum over the discrete modes $\omega_c<\omega_M$ in the cavity and an integral over the continuum extending from $0$ to $\omega_M$ outside. This sort of calculation is similar to the ones performed when dealing with the static Casimir force, and the result would similarly depends on the specific geometry considered \cite{dalvit2011casimir}. While a multimode quantum simulation is numerically infeasible in our case, its results would 
mainly boil down to a renormalization of the potential barrier felt by the atom, which can be taken into account in our model as a phenomenological renormalisation of the coupling constant $\eta_0$ without qualitatively modify our results.

\noindent \textbf{The system Hamiltonian in the dipole gauge.}~ By applying a Power-Zienau-Woolley (PZW) transformation to Eq.~\eqref{eq: total Hamiltonian (coulomb)} \cite{babiker1983derivation}, we can derive the full Hamiltonian in the dipole gauge, $\mathcal{\hat{H}}^{(\rm d)} = \mathcal{\hat{T}} \mathcal{\hat{H}}^{(\rm c)} \mathcal{\hat{T}}^\dagger $, where

\begin{eqnarray}
    \label{eq: PZW transformation matrix}
    \mathcal{\hat{T}} = \exp\left[ -i {\eta(\hat{x})} \sx (\aop + \adop) \right].
\end{eqnarray}
The transformed Hamiltonian $\mathcal{\hat{H}}^{(\rm d)}$ has two terms: the quantum Rabi Hamiltonian in dipole gauge $\mathcal{\hat{H}}_{\rm R}^{(\rm d)} = \mathcal{\hat{T}} \mathcal{\hat{H}}_{\rm R}^{(\rm c)} \mathcal{\hat{T}}^\dagger$, and the transformed kinetic contribution $\mathcal{\hat{T}} \hat{p}^2 \mathcal{\hat{T}}^\dagger / (2 m)$. The first term becomes
\begin{eqnarray}
\label{eq: rabi hamiltonian (dipole)}
\hat{\mathcal{H}}_{\rm R}^{\rm (d)} = \omega_c \adop \aop + \frac{1}{2} \omega_a \sz - i \omega_c \eta (\hat{x}) \sx (\aop - \adop) + \omega_c \eta^2 (\hat{x}), \nonumber \\
\end{eqnarray}
while the second one is
\begin{eqnarray}
    \label{eq: gauge transformation of kinetic part}
    \hat{\mathcal{T}} \hat{p}^2 \hat{\mathcal{T}}^\dag &=& \hat{p}^2 + \sx \left( \aop + \adop \right) \left[ \eta'(\hat{x}) \hat{p} + \hat{p} \eta'(\hat{x}) \right] \nonumber \\
    &+& \eta'^2 (\hat{x}) \left( \aop + \adop \right)^2,
\end{eqnarray}
this follows from the property $\hat{\mathcal{T}} \hat{p} \hat{\mathcal{T}}^\dag = \hat{p} + \eta'(\hat{x}) \sx (\aop + \adop)$, where $\eta'(x) = d\eta (x) / d x$. 

Figure \ref{fig: gauge comparison} shows the mean photon number generated when the atom passes through the cavity as a function of time for different initial kinetic energies. The figure displays both the mean value of the real photons $\langle \hat{\mathcal{X}}^{-} \hat{\mathcal{X}}^{+}\rangle$ and the mean values of the bare photon number in the Coulomb and dipole gauge.
In Figure \ref{fig: gauge comparison}a, the atom has an initial kinetic energy comparable to the bare energies of the system, ${E}_K = 1$. When it enters the cavity, the evolution is \emph{adiabatic} and \emph{no real photons are emitted}. The expectation value $\langle \hat{\mathcal{X}}^{-} \hat{\mathcal{X}}^{+}\rangle$ is exactly zero during the entire dynamics and does not depend on the chosen gauge. On the contrary, $\langle \hat a^\dagger\hat a\rangle$ is different from zero inside the cavity, and depends on the chosen gauge. A careful analysis displayed in Ref.~\cite{settineri2021gauge} shows that the Coulomb gauge mean value $\langle \hat a^\dag \hat a \rangle_c$ represents the correct mean value for bare photons. Indeed, by changing to the dipole gauge, also the photonic operators need to be transformed $\aop \to \aop^\prime = \hat{\mathcal{T}} \aop \hat{\mathcal{T}}^\dag$, obtaining in this case the same results using the Coulomb gauge $\langle \hat {a}^\prime {}^\dag \hat a^\prime \rangle_d = \langle \hat a^\dag \hat a \rangle_c$.

In Figure \ref{fig: gauge comparison}b, the atom has an initial normalised kinetic energy much higher than the bare energies of the system, ${E}_K = 30$. When it enters the cavity, the evolution of the system is \emph{non-adiabatic} and \emph{real photons are generated}. In fact, the expectation value $\langle \hat{\mathcal{X}}^{-} \hat{\mathcal{X}}^{+}\rangle$ progressively increases till the moment that the center of the atomic wave packet leaves the cavity. Instead, the mean value $\langle \hat a^\dagger\hat a\rangle_c$, in the Coulomb gauge, reaches its maximum when the center of the atomic wave-packet arrives at the center of the cavity where the coupling is the largest, and decreases afterwards. 
\begin{figure}[t]
    \centering
    \includegraphics[width = \linewidth]{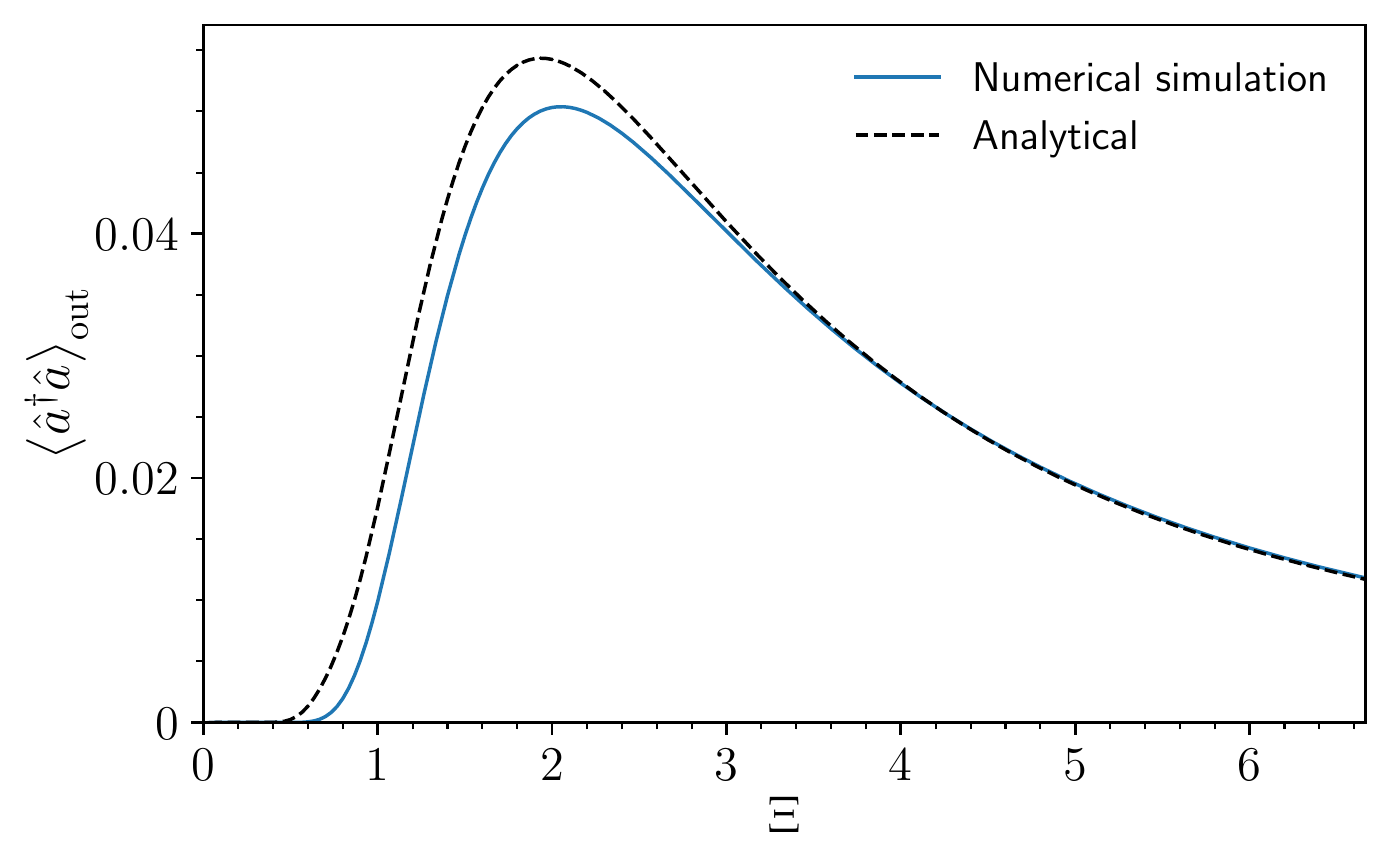}
    \caption{Comparison between numerical and analytical calculations of the output photon rate as a function of the initial normalised kinetic energy.}
    \label{fig: numerical vs analytical}
\end{figure}
\newline

\noindent\textbf{Mapping to time-dependent perturbation theory.} Using a time-dependent perturbation theory, we have analytically calculated the output photon rates.
When the initial kinetic energy is large enough, the atom velocity can be considered constant along the trajectory, because the cavity does not affect significantly the atom motion. With this assumption, we remove the motion degree of freedom and consider only the simple quantum Rabi model with an effective time-dependent coupling. In particular, the coupling follows again a Gaussian shape (but in the time domain) $\eta(t) = \eta_0 \exp [ - t^2 / 2\, \mu_t^2  ]$, with $\mu_t = \mu_c m/k_0$.
As a starting point, we expand the quantum Rabi Hamiltonian in the Coulomb gauge in Eq.\,\eqref{eq: rabi hamiltonian (coulomb)} up to the third order with respect to the photon operators. With these approximations, the Hamiltonian becomes \break $\hat{\mathcal{H}}^{\rm (c)} (t) \simeq \hat{\mathcal{H}}_0 + \hat{\mathcal{V}} (t)$, with $\hat{\mathcal{H}}_0 = \omega_c \adop \aop + (\omega_a / 2) \sz$, and
\begin{eqnarray}
    \label{eq: Rabi Hamiltonian in Coulomb (expandend)}
    \hat{\mathcal{V}} (t) &=& \frac{1}{2} \omega_a \left[ 2 \eta (t) (\aop + \adop) \sy - 2 \eta^2 (t) (\aop + \adop)^2 \sz \right. \nonumber \\
    &+& \left. \frac{8}{6} \eta^3 (t) (\aop + \adop)^3 \sy \right]~.
\end{eqnarray}
Using first-order time-dependent perturbation theory, the transition probability from the ground state $\ket{g, 0}$ to the excited states $\ket{n}$ is given by
\begin{equation*}
    c_n^{(1)} = -i \int_{- \infty}^{+ \infty} \mel{n}{\hat{\mathcal{V}} (t)}{g, 0} e^{i (\omega_n - \omega_{g, 0} ) t} dt~.
\end{equation*}
For the states $\ket{e, 1}$, $\ket{g, 2}$ and $\ket{e, 3}$, and in the case of resonant condition $\omega_a = \omega_c$, we obtain
\begin{eqnarray}
\label{eq: perturbation theory coefficients}
    c_{e, 1}^{(1)} &=& - \frac{\eta_0}{\Xi} \sqrt{2\pi}  e^{- 2 \frac{1}{\Xi^2}} + \frac{\eta_0^3}{\Xi} \sqrt{\frac{8 \pi}{3}} e^{- \frac{2}{3} \frac{1}{\Xi^2}}, \nonumber \\
    c_{g, 2}^{(1)} &=& - i \frac{\eta_0^2}{\Xi} \sqrt{2\pi} e^{- \frac{1}{\Xi^2}}, \nonumber \\
    c_{e, 3}^{(1)} &=& \frac{\eta_0^3}{\Xi} \sqrt{\frac{16 \pi}{9}} e^{- \frac{8}{3} \frac{1}{\Xi^2}}~.
\end{eqnarray}
The total photon rates can be calculated using $\langle\adop\aop\rangle$, because, at the end of the process, it predicts correct results, as shown in Fig.\,\ref{fig: gauge comparison}. We obtain,
\begin{equation}
    \label{eq: output photon number}
    \expval{\adop \aop}_{\rm out} \approx \abs{c_{e, 1}^{(1)}}^2 + 2 \abs{c_{g, 2}^{(1)}}^2 + 3 \abs{c_{e, 3}^{(1)}}^2~.
\end{equation}
Therefore, we can analytically calculate the total photon rates substituting Eqs.\,\eqref{eq: perturbation theory coefficients} in Eq.\,\eqref{eq: output photon number}. 
Figure \ref{fig: numerical vs analytical}(a) shows the comparison between analytical and numerical results of the output photon emission rate $\langle \adop \aop \rangle$ as a function of the kinetic energy. 
As expected from the approximations made to obtain the analytical formula in Eq.\,\eqref{eq: output photon number}, both numerical and analytical results coincide in the limit of high kinetic energies. Moreover, they remain in good agreement also for lower kinetic energies. We observe that the analytical calculation overestimates the generated photon rate, which is a direct consequence of neglecting the back-reaction effect on the atom motion.  Lowering the kinetic energy, as expected, the generation of photons reduces and then tends to zero. Increasing the kinetic energy increases the photon rate untill a maximum value, after that the photon rate decreases because the atom passes trough the cavity faster than the time required to generate excitations.

\bibliography{biblio}
\section*{Acknowledgements}
S.S. acknowledges the Army Research Office (ARO) (Grant No. W911NF-19-1-0065).
S.D.L. is supported by a Royal Society Research fellowship, the Philip Leverhulme prize, and he acknowledges support from the Leverhulme Grant No. RPG-2022-037.
F.N. is supported in part by:
Nippon Telegraph and Telephone Corporation (NTT) Research,
the Japan Science and Technology Agency (JST) [via
the Quantum Leap Flagship Program (Q-LEAP), and
the Moonshot R\&D Grant Number JPMJMS2061],
the Japan Society for the Promotion of Science (JSPS)
[via the Grants-in-Aid for Scientific Research (KAKENHI) Grant No. JP20H00134],
the Army Research Office (ARO) (Grant No. W911NF-18-1-0358),
the Asian Office of Aerospace Research and Development (AOARD) (via Grant No. FA2386-20-1-4069), and
the Foundational Questions Institute Fund (FQXi) via Grant No. FQXi-IAF19-06.
\end{document}